\documentclass{appolb}
\usepackage{epsfig}

\begin{document}
\title{Prompt Photon Production at the Tevatron%
\thanks{Presented at the PHOTON2005 conference, September 2005,
Warsaw, Poland}%
}
\author{STEFAN S\"OLDNER-REMBOLD
\address{University of Manchester\\
School of Physics and Astronomy\\
Manchester M13 9PL\\
United Kingdom\\
\bigskip
on behalf of the D\O\ and CDF Collaborations}}
\maketitle
\begin{abstract}
The D\O\ and CDF experiments have measured prompt photon production
using Run II data taken at a centre-of-mass energy $\sqrt{s}$ of 
$1.96$~TeV. The results are compared to different types of perturbative
QCD calculations.
\end{abstract}
  
\section{Introduction}
In Leading Order (LO) 
prompt photons originating in the hard interaction between two partons are
produced mainly via quark-gluon Compton
scattering ($qg\to q\gamma$) or quark--anti-quark
annihilation ($q\bar{q}\to g\gamma$). 
Studies of direct photons with large transverse momenta, $p_T^\gamma$,
provide precision tests of perturbative QCD (pQCD) as well as
information on the distribution of partons within protons,
particularly the gluon.  These data are also used in global fits of parton
distributions functions (PDFs).  

The LO contributions to di-photon production are
quark--anti-quark annihilation ($q\bar{q}\to\gamma\gamma$)
and gluon-gluon scattering ($gg\to\gamma\gamma$). The latter
subprocess involves initial state gluons coupling to the photons
through a quark box; thus the subprocess is suppressed by a
factor $\alpha^2_s$. The rate is still high for small $\gamma\gamma$
masses due to the large flux of gluons. 
Processes where both photons originate from
parton fragmentation or where one photon is prompt and one
photon is from parton fragmentation also contribute in LO.
Di-photon final states are not only interesting to study pQCD
but they are also signatures for many new physics processes, such
as Higgs production at the LHC or Large Extra Dimensions.

D\O\ has measured the cross section for the
inclusive production of isolated photons in the range
$23<p_T^\gamma<300$ GeV.  
This extends previous measurements in this energy
regime~[1--5] to significantly higher values of $p_T^\gamma$.  
CDF has measured the di-photon cross-section in
$p\bar{p}$ collisions~\cite{CDFpaper}.
Both measurements are restricted to photons in the pseudorapidity range
$|\eta|<0.9$. 
The data samples correspond to an integrated luminosity of about
$L=326$~pb$^{-1}$ for D\O\ and $L=107$~pb$^{-1}$ for CDF.

Photons from energetic $\pi^0$ and $\eta$
mesons are the main background to direct photon production especially
at small $p_T^\gamma$. Since these mesons
are produced inside jets, their contribution is suppressed with
respect to direct photons by requiring the photon be isolated from
other particles. 

\section{Prompt Inclusive Photon Production (D\O)}

Photon candidates in D\O\ were formed from clusters of calorimeter cells
within a cone. Candidates
were selected if there was significant energy in the electromagnetic
(EM) calorimeter
layers ($>95\%$), and the probability to have a spatially-matched track was 
less than $0.1\%$, and they satisfied an isolation requirement.
Potential backgrounds from cosmic rays and
leptonic $W$ boson decays were suppressed by requiring the missing
transverse energy to be less than $0.7p_T^\gamma$.  
Four additional variables were input to an artificial neural network
(NN) to further suppress
background and to estimate the purity of the resulting photon sample.
The NN was trained to discriminate between direct photons and
QCD as well as electroweak background events. The total
number of photon candidates remaining after these requirements
is 2.7 million.

The isolated-photon cross section $d^2\sigma/(dp_Td\eta)$
is measured by performing an unsmearing as a function of
$p_T^\gamma$. This is done by iteratively fitting the convolution of an ansatz
function with an energy resolution function.  The uncertainty in this
correction was estimated using two different ansatz functions and
included the uncertainty in the energy resolution.  An additional
correction was applied to $p_T^\gamma$ for the difference in the
energy deposited in the material upstream of the calorimeter between
electrons and photons.  

\begin{figure}
\centering
\includegraphics[scale=0.33,bb=5 0 515 495,clip=true]{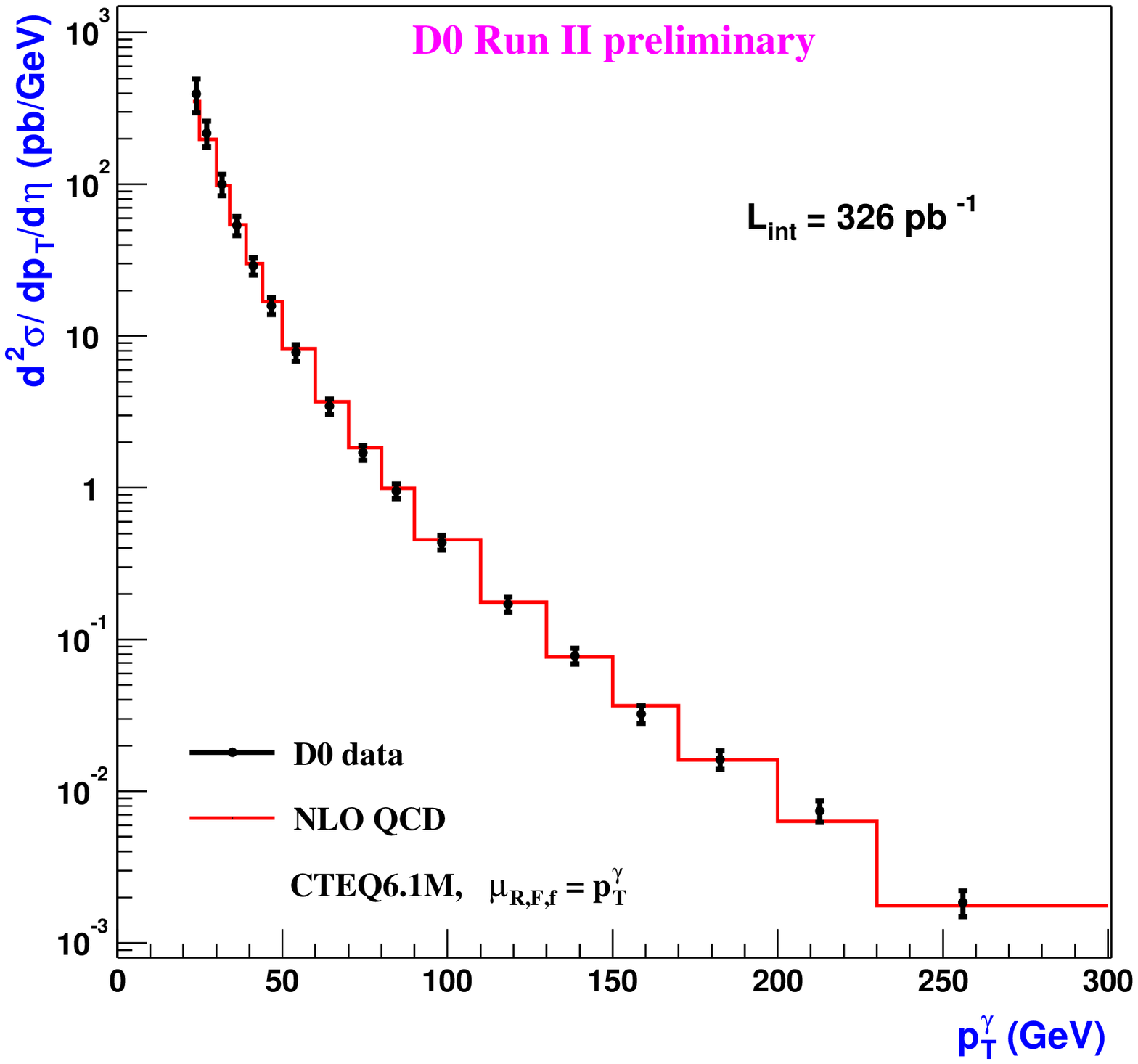}
\includegraphics[scale=0.33,bb=5 0 515 495,clip=true]{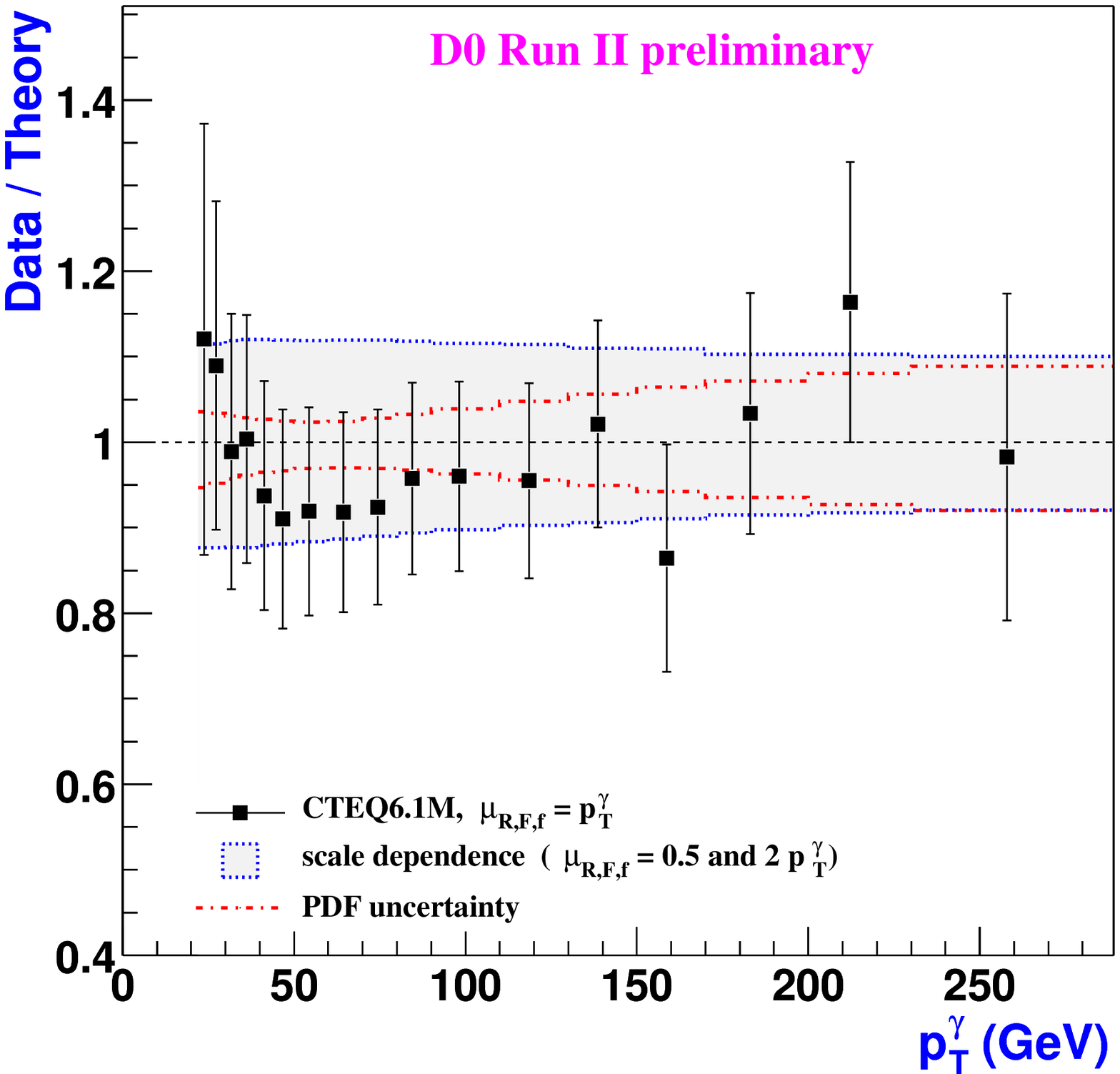}
\caption{
a) The inclusive cross section for the production of isolated photons as
a function of $p_T^\gamma$.  The results from the NLO pQCD calculation
with {\sc jetphox} are shown as solid line.
b) The ratio of the measured
cross section to the theoretical predictions from {\sc jetphox}. The
full vertical lines correspond to the overall uncertainty while the
internal line indicates just the statistical uncertainty. Dashed lines
represents the change in the cross section when varying the
theoretical scales by factors of two.  The shaded region indicates the
uncertainty in the cross section estimated with CTEQ6.1 PDFs.
\label{fig:DTleft}}
\end{figure}

The measured cross section, together with statistical and systematic
uncertainties, is presented in Fig.~\ref{fig:DTleft}a. Sources of
systematic uncertainty include luminosity ($6.5\%$), event vertex
determination ($3.6\%-5.0\%$), energy calibration ($9.6\%-5.5\%$), the
fragmentation model ($7.3\%-1.0\%$), photon conversions ($3\%$), and
the photon purity fit uncertainty as
well as statistical uncertainties on the determination of geometrical
acceptance ($1.5\%$), trigger efficiency ($11\%-1\%$), selection
efficiency ($5.4\%-3.8\%$) and unsmearing ($1.5\%$). 
The uncertainty ranges are quoted for increasing $p_T^{\gamma}$.
Most of the
systematic uncertainties have large ($>80\%$) bin-to-bin correlations
in $p_T^\gamma$.

Results from a next-to-leading order (NLO) pQCD calculation ({\sc
jetphox}~\cite{DIPHOX,JETPHOX}) are compared to the measured D\O\ cross
section in Fig.~\ref{fig:DTleft}a.  These results were derived using
the CTEQ6.1M~\cite{cteq61} PDFs and the BFG~\cite{BFG} fragmentation
functions (FFs).  The renormalization, factorization, and
fragmentation scales were chosen to be
$\mu_{R}\!=\!\mu_{F}\!=\!\mu_{f}\!=\!p_T^\gamma$. 
As shown in Fig.~\ref{fig:DTleft}b, the calculation agrees,
within uncertainties, with the measured cross section.  The scale
dependence, estimated by varying scales by
factors of two, are displayed in Fig.~\ref{fig:DTleft}b as dashed
lines.  The span of these results is comparable to the overall
uncertainty in the cross section measurement.  The filled area 
represents the uncertainty associated with the
CTEQ6.1M PDFs.  The central values of the predictions changes by less
than $7\%$ when the PDF is replaced by MRST2004~\cite{Martin:2004ir}
or Alekhin2004~\cite{Alekhin:2002fv}.  The calculation is also
sensitive to the implementation of the isolation requirements
including the hadronic fraction in the ${\mathcal R}=0.2$ cone around
the photon at a level of $3\%$.

\begin{figure}
\centering
\includegraphics[scale=0.31]{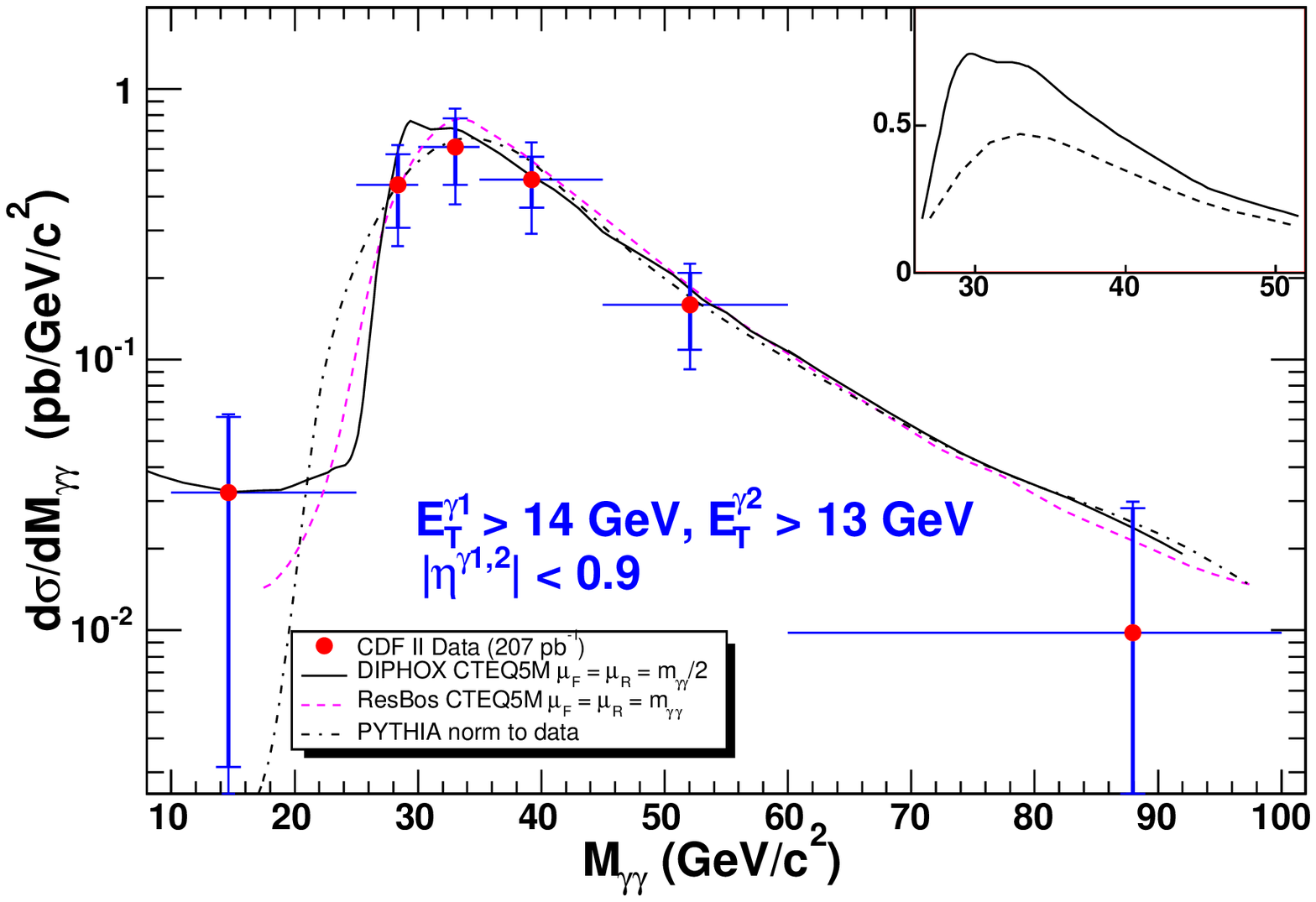}
\includegraphics[scale=0.31]{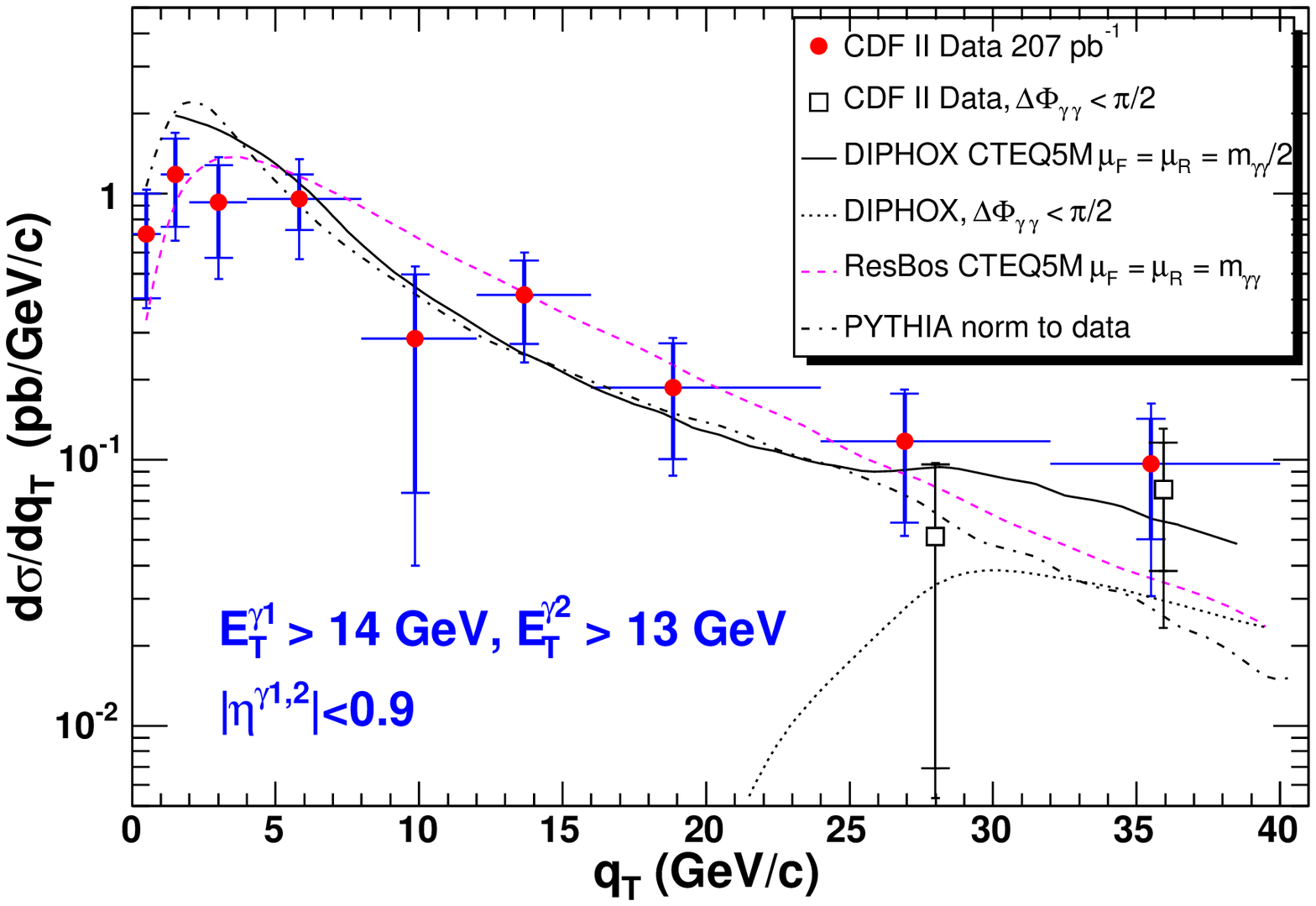}\\
\includegraphics[scale=0.315]{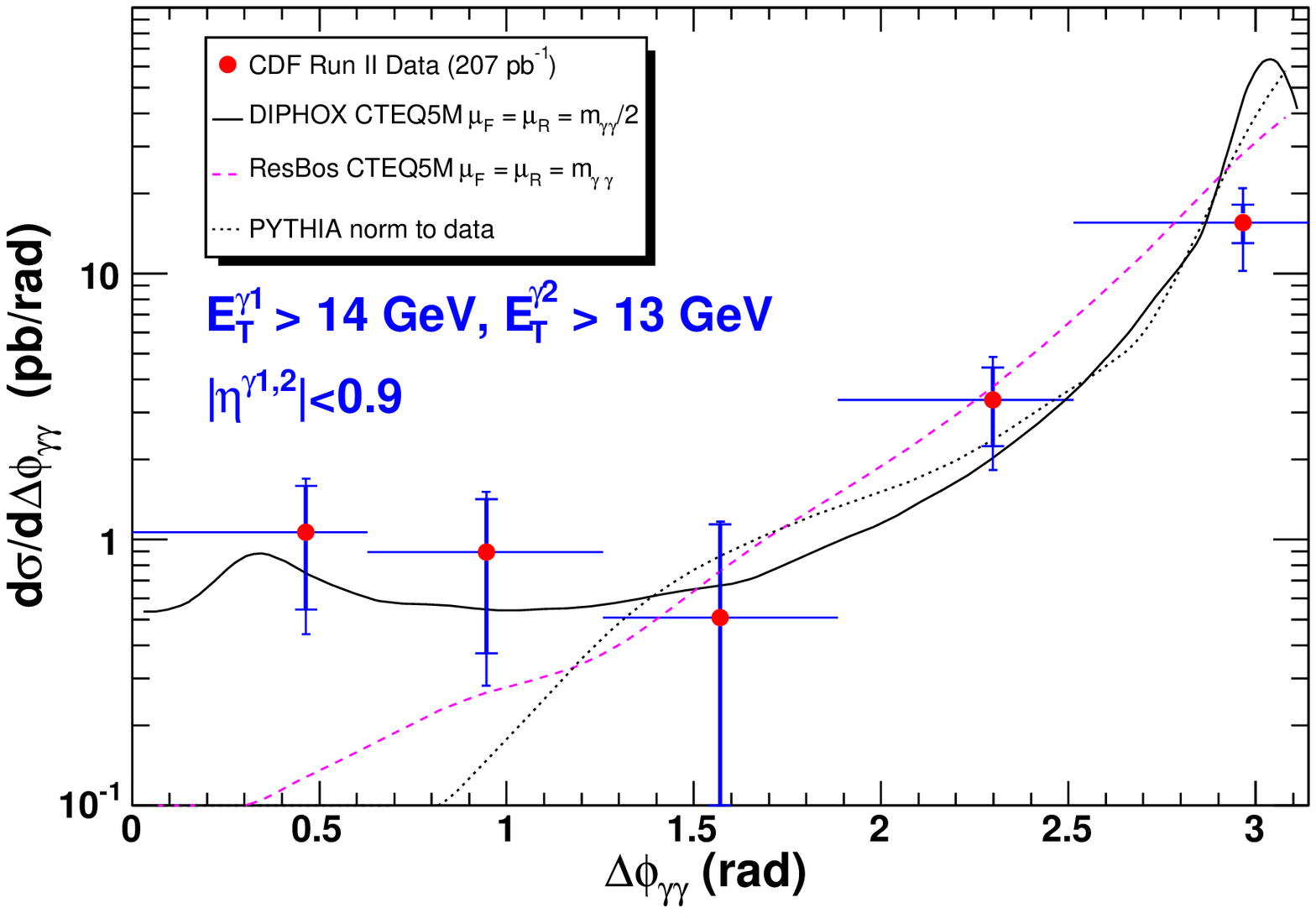}
\caption{\label{cdf}
a) The di-photon differential cross section measured by CDF as function of the 
invariant mass.
The inset shows, on a linear scale, the total NLO cross section from 
{\sc diphox} with (solid) and without (dashed) the gluon-gluon contribution.
b)  The differential cross section as function of the di-photon system 
$p_T$ (referred as "$q_T$") 
Also shown, at larger 
$q_T$, are the {\sc diphox} prediction (dot) and the CDF data (open squares) 
for the configuration where the two photons are required to have 
$\Delta\phi < \pi/2$.
c) The differential cross section measured by CDF as function of $\Delta\phi$ 
between the two photons, along with predictions from {\sc diphox} (solid).
The predictions from {\sc diphox} (solid), 
{\sc ResBos} (dashed), and {\sc pythia} (dot-dashed) are also shown.
The {\sc pythia} predictions have been scaled up by a factor of two. 
}
\end{figure}

\section{Prompt Di-Photon Production (CDF)}

Photon candidates in CDF were identified by requiring the ratio
of the hadronic to EM energy to be less than $0.055+0.00045E$, where
$E$ is the EM energy. Photon candidates with any associated tracks with
$p_T>0.5$~GeV  were rejected and the lateral profile of EM
showers in the calorimeter is compared to the profile of electrons
measured in a test beam. After the final selection, $889$ di-photon
events remain, of which $427\pm59$~(stat)  
are real $\gamma\gamma$ events.
This background from neutral mesons
such as $\pi^{0}$ and $\eta$ is determined in each
kinematic bin using shower shape variables and hits
in the preshower detector.

From these events, 
the calculated acceptance and the integrated luminosity, 
CDF has determined the di-photon 
cross sections for several kinematic variables.
The $\gamma\gamma$ mass distribution is shown in Fig.~\ref{cdf}a,
along with NLO predictions from {\sc diphox}~\cite{DIPHOX} and 
{\sc ResBos}~\cite{ResBos} and from
the LO Monte Carlo {\sc pythia}~\cite{PYTHIA}. 
{\sc diphox} is a fixed-order NLO QCD calculation.
{\sc ResBos} resums the effects of initial state soft gluon
 radiation. This  is particularly important for 
the distribution of the transverse momentum of the
di-photon system, $q_T$, which is a delta function at LO
and divergent as $q_T\rightarrow0$ at NLO. 
The $q_T$ distribution is shown in Fig.~\ref{cdf}b, 
and the $\Delta\phi$ distribution between the two photons is shown in 
Fig.~\ref{cdf}c. The systematic effects include uncertainties on
 the selection efficiencies ($11\%$), uncertainties 
from the background subtraction ($20-30\%$)  and from
 the luminosity determination ($6\%$). 

The observed differences between the predictions are expected. 
The {\sc ResBos} $q_T$ prediction is smooth in the entire range, 
while the  {\sc diphox} curve is unstable at low $q_{T}$ due
to the NLO singularity.
The fragmentation contribution in {\sc ResBos} is effectively at LO. Since 
fragmentation to a photon is of order $\alpha_{em}/\alpha_{s}$,
some 2$\rightarrow$3 processes such as $qg\rightarrow gq\gamma$,
where the quark in the final state fragments to a second photon, are 
of order $\alpha_{em}^{2}\alpha_{s}$  and are included in a full NLO
calculation. These contributions are present in {\sc diphox}, but not in {\sc ResBos},
which leads to an underestimation of the production rate in {\sc ResBos}  
at high $q_{T}$, low $\Delta\phi$, and low $\gamma\gamma$ mass.  
In particular, the shoulder at $q_T\approx 30$~GeV
arises from an increase in phase space for both the direct and fragmentation
 subprocesses~\cite{DIPHOXFrag}. 
The $q_T$ prediction for the  
$\Delta\phi < \pi/2$ region in Fig.~\ref{cdf}b demonstrates
that the bump in the {\sc diphox} prediction at
 a $q_T\approx 30$~GeV is due  to the ``turn-on'' 
of the $\Delta\phi<\pi/2$ region of phase space. At $\Delta\phi$ values
 above $\pi/2$, the effects from soft gluon emission (included in {\sc ResBos}
 but not in {\sc diphox}) are significant.

The data are in good agreement with the predictions for the mass distribution. 
At low to moderate $q_{T}$ and $\Delta\phi$ 
greater than $\pi/2$, where the effect of soft 
gluon emissions are important, the data agree better with {\sc ResBos} than {\sc diphox}.
 By contrast, in the regions where the 2$\rightarrow$3  
fragmentation contribution becomes important, \ie large $q_{T}$,  $\Delta\phi$ 
less than $\pi/2$ and low di-photon mass, the data agree better with {\sc diphox}.

\section{Summary}
CDF and D\O\ have measured prompt photon production using
the Run II data taken at the Tevatron with data samples more twice
the size of the Run I data. In general, predictions of NLO pQCD are 
in good agreement with the data in different regions of phase space.

\section*{Acknowledgement}
Special thanks to Dmitry Bandurin and Michael Begel for their
help in preparing these proceedings.


\begin{thebibliography}{99}


\bibitem{Abe:1994rr}
F. Abe et al., CDF Collab., Phys. Rev. Lett. {\bf 73}, 2662 (1994).
\bibitem{Abbott:1999kd}
B. Abbott et al., D\O\ Collab., Phys. Rev. Lett. {\bf 84}, 2786 (2000).
\bibitem{Abazov:2001af}
V.M. Abazov et al., D\O\ Collab., Phys. Rev. Lett. {\bf 87}, 251805 (2001).
\bibitem{Acosta:2002ya}
D. Acosta et al., CDF Collab., Phys. Rev. D{\bf 65}, 112003 (2002).
\bibitem{Acosta:2004bg}
D. Acosta et al., CDF Collab., Phys. Rev. D{\bf 70}, 074008 (2004).
\bibitem{CDFpaper}
D. Acosta et al., CDF Collab., Phys. Rev. Lett. {\bf 95}, 022003 (2005).
\bibitem{DIPHOX} T.~Binoth et al.,
Eur.\ Phys.\ J.\ C{\bf 16}, 311 (2000).
\bibitem{JETPHOX}
S. Catani et al., JHEP {\bf 05}, 028 (2002).
\bibitem{cteq61}
D. Stump et al., JHEP {\bf 10}, 046 (2003).
\bibitem{BFG}
L. Bourhis et al., Eur.\ Phys.\ J.\ C{\bf 2}, 529 (1998).
\bibitem{Martin:2004ir}
A.D. Martin et al., Phys. Lett. B{\bf 604}, 61 (2004).
\bibitem{Alekhin:2002fv}
S. Alekhin, Phys. Rev. D{\bf 68}, 014002 (2003).
\bibitem{ResBos} C.~Balazs et al.,
Phys.\ Rev.\ D{\bf 57}, 6934 (1998).
\bibitem{PYTHIA} T. Sj\"ostrand et al.,
Comp. Phys. Commun. {\bf 135 } 238 (2001).
\bibitem{DIPHOXFrag} T.~Binoth et al.,, 
Phys.\ Rev.\ D{\bf 63}, 114016 (2003).
\end{thebibliography}
\end{document}